\documentclass[sigconf]{acmart}
\pdfoutput=1

\usepackage{booktabs} 
\usepackage{amsmath}
\usepackage{amsthm}
\usepackage{url}

\usepackage{adjustbox}

\usepackage{booktabs}
\usepackage{multirow,multicol}

\usepackage{tikz,pgfplots}
\pgfplotsset{compat=1.13}
\usepackage{pgfplotstable}
\usepgfplotslibrary{groupplots}
\usetikzlibrary{arrows,decorations.markings}
\usetikzlibrary{shapes.arrows}
\usetikzlibrary{fit}
\usetikzlibrary{decorations.pathmorphing,calc,shapes,shapes.geometric,patterns}

\usetikzlibrary{fadings}
\tikzfading[name=fade l,left color=transparent!20,right color=transparent!100]
\tikzfading[name=fade r,right color=transparent!20,left color=transparent!100]
\tikzfading[name=fade d,bottom color=transparent!20,top color=transparent!100]
\tikzfading[name=fade u,top color=transparent!20,bottom color=transparent!100]

\tikzset{textnode/.style={inner sep=0pt,outer sep=0,execute at begin node={\strut}}}
\tikzstyle{opencircle} = [circle,minimum width=10, draw, inner sep=1.5]
\tikzstyle{itxset} = [rounded corners=3pt, draw, minimum height=14pt, inner sep=0]
\tikzstyle{internalnode} = [textnode,circle, draw, fill, text=white]
\tikzstyle{externalnode} = [textnode, circle, draw]
\tikzstyle{nonterminal} = [textnode,draw,inner xsep=1.5]
\tikzstyle{itxsetfocus} = [itxset, ultra thick]

\selectcolormodel{cmyk}

\newcommand{\etal}{\textit{et~al.}}
\newcommand{\eg}{\textit{e.g.}}
\newcommand{\ie}{\textit{i.e.}}
\newcommand{\nop}[1]{}

\newcommand\framenode[2][10pt]{
    \fill[white,path fading=fade u] ($(#2.south west)+(0,0.25pt)$) rectangle ($(#2.south east)-(0, #1)$);
    \fill[white,path fading=fade d] ($(#2.north west)-(0,0.25pt)$) rectangle ($(#2.north east)-(0,-#1)$);
    \fill[white,path fading=fade l] ($(#2.south east)+(-0.25pt,0)$) rectangle ($(#2.north east)-(-#1,0)$);
    \fill[white,path fading=fade r] ($(#2.south west)+(0.25pt,0)$) rectangle ($(#2.north west)-( #1,0)$);
}

\newenvironment{customlegend}[1][]{%
    \begingroup
    \csname pgfplots@init@cleared@structures\endcsname
    \pgfplotsset{#1}%
}{%
    \csname pgfplots@createlegend\endcsname
    \endgroup
}%
\def\addlegendimage{\csname pgfplots@addlegendimage\endcsname}
\newcolumntype{C}{>{\raggedleft\arraybackslash}p{3mm}}

\newcommand{\n}[1]{\ensuremath{\text{N}^{\rule{0pt}{6pt}#1}}}
\newcommand{\nss}[2]{\ensuremath{\text{N}^{\rule{0pt}{6pt}#1}_{\rule[-2pt]{0pt}{0pt}{#2}}}}

\acmConference[KDD'18]{KDD}{August 2018}{London, UK}

\begin{document}
\title{Growing Better Graphs
With Latent-Variable Probabilistic
Graph Grammars}

\author{Xinyi Wang}
\affiliation{%
  \institution{Carnegie Mellon University}
  \city{Pittsburgh}
  \state{PA}
  \postcode{15213}
}
\email{xinyiw1@cs.cmu.edu}

\author{Salvador Aguinaga}
\affiliation{%
  \institution{University of Notre Dame}
  \city{Notre Dame}
  \state{IN}
  \postcode{46556}
}
\email{saguinag@nd.edu}

\author{Tim Weninger}
\affiliation{%
  \institution{University of Notre Dame}
  \city{Notre Dame}
  \state{IN}
  \postcode{46556}
}
\email{tweninge@nd.edu}

\author{David Chiang}
\affiliation{%
  \institution{University of Notre Dame}
  \city{Notre Dame}
  \state{IN}
  \postcode{46556}
}
\email{dchiang@nd.edu}

\begin{abstract}
Recent work in graph models has found that probabilistic hyperedge replacement grammars (HRGs) can be extracted from graphs and used to generate new random graphs with graph properties and substructures close to the original. In this paper, we show how to add latent variables to the model, trained using Expectation-Maximization, to generate still better graphs,  that is, ones that generalize better to the test data. We evaluate the new method by separating training and test graphs, building the model on the former and measuring the likelihood of the latter, as a more stringent test of how well the model can generalize to new graphs. On this metric, we find that our latent-variable HRGs consistently outperform several existing graph models and provide interesting insights into the building blocks of real world networks.
\end{abstract}

\maketitle


\section{Introduction}
Detecting, interpreting and comparing structures and properties of network data about social interactions and complex physical phenomena is critically important to a variety of problems. However, this is a difficult task because comparisons between two or more networks can involve checking for graph or subgraph isomorphism, for which no tractable solution is known. Instead, various network properties (\eg, degree distribution, centrality distributions) have been used to describe and compare networks. 

Another approach is to consider a network's global structure as a by-product of a graph's local substructures~\cite{ugander2013subgraph}. More sophisticated graph statistics are based on counting the number of small motifs \cite{milo2002network} or graphlets~\cite{ahmed2015efficient} present in the graph and comparing their distributions \cite{yaveroglu2014revealing}. Unfortunately, graphlet counting presupposes that all possible graphlets be enumerated ahead of time. Because the number of unique graphlets increases exponentially with the number of nodes in the graphlet, previous work has been limited to graphlets of at most five nodes.

An alternative to developing sophisticated graph statistics is to learn graph generation models that encode properties of the graph in various ways. Graph generators like the Exponential Random Graph Model (ERGM) \cite{robins+al}, the Chung-Lu Edge Configuration Model (CL) \cite{chung+lu}, the Stochastic Kronecker Graph (SKG) \cite{leskovec2010kronecker}, and the Block Two-Level Erd\H{o}s-R\'{e}nyi Model (BTER) \cite{seshadhri2012community} can be fitted to real-world graphs. 

Recent work has found that many social and information networks have a more-or-less tree-like structure, which implies that detailed topological properties can be identified and extracted from a graph's {\em tree decomposition}~\cite{adcock2016tree}. 
Based on these findings, Aguinaga~\etal~described a method to turn a graph's tree decomposition into a \emph{Hyperedge Replacement Grammar} (HRG) 
~\cite{aguinaga2016growing}. The HRG model can then generate new graphs with properties similar to the original. 

One limitation of the HRG model is that the instructions for reassembling the building blocks, \ie, the graph grammar, encode only enough information to ensure that the result is well-formed. HRG production rules are extracted directly from the tree decomposition; some rules from the top of the tree, some from the middle of the tree, and some from the bottom of the tree. Then, to generate an new graph that is similar to the original graph, we would expect that rules from the top of the tree decomposition are applied first, rules from the middle next, and rules from the leaves of the tree are applied last. However, when generating a new graph the HRG model applies rules probabilistically; where a rule's probability relative to its frequency in the grammar.  However, when generating a new graph, the rules in HRG models have no context on when they should fire. HRG models are in need of a mechanism that corrects for this problem by providing context to the rules.

In the present work, we make three contributions:
\begin{enumerate}
    \item We improve the HRG model by encoding context in latent variables. 
    \item We propose a methodology for evaluating our model that enforces a strict separation between training and test data, in order to guard against overfitting.
    \item We test our model on 6 train/test pairs of graphs and find that it discovers a better model, that is, one that generalizes better to the test data, than the original HRG model as well as Kronecker and Chung-Lu models.
\end{enumerate}

\section{Background}

Before we introduce our model, we first provide an overview and examples of the HRG model.

\subsection{Hyperedge Replacement Grammars}

Like a context free string grammar (CFG), an HRG has a set of production rules $A \rightarrow R$, where $A$ is called the left-hand side (LHS) and $R$ is called the right-hand side (RHS). 
In an HRG, a rule's RHS is a graph (or hypergraph) with zero or more \emph{external} nodes. Applying the rule replaces a hyperedge labeled $A$ with the graph $R$; the nodes formerly joined by the hyperedge are merged with the external nodes of $R$. The HRG generates a graph by starting with the start nonterminal, $S$, and applying rules until no more nonterminal-labeled hyperedges remain.

\subsection{Tree Decomposition}
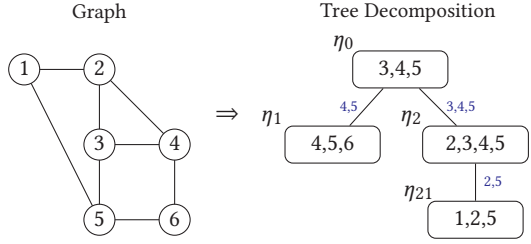
\begin{figure}
\centering
\begin{tikzpicture}

\node (s) at (-0.7,2.25) {\small Graph};
\node [opencircle,textnode] (v1) at (-1.7,1.5) {1};
\node [opencircle,textnode] (v2) at (-0.7,1.5) {2};
\node [opencircle,textnode] (v3) at (-0.7,0.5) {3};
\node [opencircle,textnode] (v4) at (0.3,0.5) {4};
\node [opencircle,textnode] (v5) at (-0.7,-0.5) {5};
\node [opencircle,textnode] (v6) at (0.3,-0.5) {6};
\draw  (v1) edge (v2);
\draw  (v1) edge (v5);
\draw  (v2) edge (v3);
\draw  (v2) edge (v4);
\draw  (v3) edge (v4);
\draw  (v3) edge (v5);
\draw  (v4) edge (v6);
\draw  (v5) edge (v6);

\node [] (e) at (2.55,1.85) {$\eta_0$};
\node [] (e) at (1.6,0.85) {$\eta_1$};
\node [] (e) at (3.45,0.85) {$\eta_2$};
\node [] (e) at (3.55,-0.15) {$\eta_{21}$};

\node (s) at (3.4,2.25) {\small Tree Decomposition};

\node [minimum width=25, minimum height=25](mc) at (1.0,.9) {$\Rightarrow$};

\node [itxset,minimum width=36] (x) at (3.3,1.5) {3,4,5};
\node [itxset,minimum width=40] (y) at (4.3,0.5) {2,3,4,5};

\draw (x) edge node [blue, midway, right] { \scriptsize{3,4,5} } (y);

\node[itxset,minimum width=36](z) at (4.3,-0.5) {1,2,5};
\draw (y) edge node [blue, midway, right] { \scriptsize{2,5} } (z);

\node [itxset,minimum width=36](w) at (2.4,0.5) {4,5,6};
\draw (x) edge node [blue, midway, left] { \scriptsize{4,5} } (w);

\end{tikzpicture}
\caption{An example graph and its tree decomposition. The width of this tree decomposition is 3, \ie, the size of the largest bag minus 1. The sepset between each bag and its parent is labeled in blue. Bags are labeled ($\eta_0$, etc.) for illustration purposes only. 
}
\label{fig:td}
\end{figure}

Given a graph $H = (V,E)$, a \emph{tree decomposition} is a tree whose nodes, called \emph{bags}, are labeled with subsets of $V$, in such a way that the following properties are satisfied:
\begin{itemize}
\item For each node $v \in V$, there is a bag $\eta$ that contains $v$.
\item For each edge $(u,v) \in E$, there is a bag $\eta$ that contains $u$ and $v$.
\item If bags $\eta$ and $\eta^\prime$ contain $v$, then all the bags on the path from $\eta$ to $\eta^\prime$ also contain $v$.
\end{itemize}

If $\eta^\prime$ is the parent of $\eta$, define $\bar{\eta} = \eta' \cap \eta$ (also known as the \emph{sepset} between $\eta'$ and $\eta$). If $\eta$ is the root, then $\bar{\eta} = \emptyset$.

All graphs can be decomposed (though not uniquely) into a tree decomposition, as shown in Fig.~\ref{fig:td}. In simple terms, a tree decomposition of a graph organizes its nodes into overlapping \emph{bags} that form a tree. The \emph{width} of the tree decomposition, which is related to the size of the largest bag, measures how tree-like the graph is. Finding optimal tree decompositions is NP-hard, but there is significant interest in finding fast approximations because many computationally difficult problems can be solved efficiently when the data is constrained to be a tree-like structure. 

\subsection{Grammar Extraction}
\label{sec:extraction}

\begin{figure}
\centering
\begin{tikzpicture}


\node (s) at (3.4,2.25) {\small Tree Decomposition};

\node [] (e) at (2.55,1.85) {$\eta_0$};
\node [] (e) at (1.6,0.85) {$\eta_1$};
\node [] (e) at (3.45,0.85) {\textbf{$\eta_2$}};
\node [] (e) at (3.55,-0.15) {$\eta_{21}$};

\node [itxset,minimum width=36] (x) at (3.3,1.5) {3,4,5};
\node [itxsetfocus,minimum width=40] (y) at (4.3,0.5) {\textbf{2,3,4,5}};

\draw (x) edge node [blue, midway, right] { \scriptsize{\textbf{3,4,5}} } (y);

\node[itxset,minimum width=36](z) at (4.3,-0.5) {1,2,5};
\draw (y) edge node [blue, midway, right] { \scriptsize{\textbf{2,5}} } (z);

\node [itxset,minimum width=36](w) at (2.4,0.5) {4,5,6};
\draw (x) edge node [blue, midway, left] { \scriptsize{4,5} } (w);

\begin{scope}[xshift=-0.5in]
\node (s) at (8,2.25) {\small Rule};

\node at (7.75,0.5) {$\n3 \rightarrow {}$};

\node [internalnode] (a) at (9.25,0.5) {3};
\node [internalnode] (b) at (10,0.5) {4};
\node [internalnode] (c) at (9.25,-0.5) {5};
\node [externalnode] (x) at (9.25,1.5) {2};

\draw (x) -- (a);
\draw (x) -- (b);

\node [nonterminal] (n1) at (8.5,1.5) {\n2};

\draw (x) edge (n1);
\draw (c) edge (n1);
\end{scope}

\end{tikzpicture}
\caption{Extraction of an HRG production rule from $\eta_2$ containing graph vertices \{2,3,4,5\}. The LHS of the production rule corresponds to the sepset of the bag and its parent. The RHS of the production rule contains nodes from the bag, terminal edges induced from the original graph, and nonterminal edges from the sepset between the bag and its children.}  
\label{fig:rule}
\end{figure}
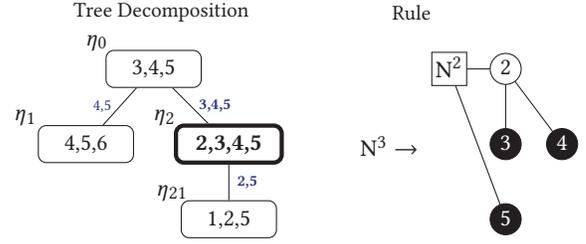

Aguinaga et al.~\cite{aguinaga2016growing} extract HRG rules from a graph, guided by a tree decomposition of the graph. For example, Figure~\ref{fig:rule} illustrates how one HRG rule is extracted from a tree decomposition. 

If we assume that the tree decomposition is rooted, then every bag $\eta$ of the tree decomposition corresponds to an edge-induced subgraph, which we write $G_\eta$, defined as follows: For each edge $(u,v) \in E$, if every bag $\eta'$ containing $u,v$ is either equal to $\eta$ or a descendant of $\eta$, then $(u,v) \in H_\eta$.
For example, in Figure~\ref{fig:td}, the bag $\eta_2 = \{2,3,4,5\}$ corresponds to the subgraph $H_{\eta_2}$ whose edges are 1--2, 1--5, 2--3, 2--4, and 3--5.

If $H=(V,E)$ is a graph and $H^\prime=(V^\prime,E^\prime)$ is an edge-induced subgraph of $H$, we define an \emph{external} node of $H^\prime$ to be any node of $H^\prime$ that has a neighbor not in $H^\prime$. Then, define the operation of \emph{replacing} $H^\prime$ with a hyperedge to be:
\begin{itemize}
\item Remove all edges in $E'$.
\item Remove all nodes in $V'$ except for the external nodes.
\item Add a hyperedge joining the external nodes.
\end{itemize}

Every bag $\eta$ also induces a HRG rule $\text{N}^{|\bar{\eta}|} \rightarrow R$, where $R$ is constructed as follows. 
\begin{itemize}
\item Make a copy of $H_\eta$.
\item Label the nodes in $\bar{\eta}$ as external nodes.
\item For each child $\eta_i$ of $\eta$, replace $H_{\eta_i}$ with a hyperedge labeled $\text{N}^{|\bar{\eta}_i|}$.
\end{itemize}
For example, in Figure~\ref{fig:rule}, the bag $\eta_2$ induces the rule shown at right. The LHS is $\text{N}^3$ because the sepset between $\eta_2$ and its parent has three nodes ($3,4,5$); in the RHS, these three nodes are marked as external. The node numbers are for illustration purposes only; they are not actually stored with the production rules. The RHS has two terminal edges (2--3, 2--4) from the original graph and one nonterminal edge (2--5) corresponding to the sepset between $\eta_2$ and its one child. 

After an HRG is extracted from the tree decomposition, its production rules are gathered into a set, merging identical rules and assigning to each unique rule $(A\rightarrow R)$ a probability $P(A \rightarrow R) = P(R \mid A)$ proportional to how many times it was encountered during extraction. This grammar can then be used to randomly generate new graphs, or compute the probability of another graph.

\section{Latent Variable Probabilistic Hyperedge Replacement Grammars}


Here, we improve upon the HRG model by encoding more context into the model via latent variables, in a process that is analogous to how a first-order Markov chain can simulate a higher-order Markov chain by expanding the state space.

In this section, we adopt a notational shortcut. In an HRG production $A \rightarrow R$, the RHS $R$ is a hypergraph fragment containing zero or more hyperedges with nonterminal labels $Y_1, \ldots, Y_r$. We suppress the graph structure of $R$ and write the rule simply as $X \rightarrow Y_1 \cdots Y_r$. 

\subsection{Nonterminal Splitting}

Following previous work on probabilistic CFGs \cite{matsuzaki+al:2005,petrov+al:2006}, we increase the context-sensitivity of the grammar by splitting each nonterminal symbol $X$ in the grammar into $n$ different \emph{subsymbols}, $X_i, \ldots, X_n$, which could potentially represent different contexts that the rule is used in. Thus, each rule in the original grammar is replaced with several \emph{subrules} that use all possible combinations of the subsymbols of its nonterminal symbols. 

For example, if $n=2$, the rule $\n2\rightarrow \epsilon$ would be split into $\n2_1\rightarrow \epsilon$ and $\n2_2\rightarrow \epsilon$.

In general, a rule with $r$ nonterminal symbols on its right-hand side is split into $n^{r+1}$ subrules.

\subsection{Learning}

\newcommand{\inside}[2]{\ensuremath{P_{\textit{in}}( #1,#2)}}
\newcommand{\outside}[2]{\ensuremath{P_{\textit{out}}(#1,#2)}}
\newcommand{\ruleprob}[1]{\ensuremath{P(#1)}}

After obtaining an $n$-split grammar from the training graphs, we want to learn probabilities for the subrules that maximize the likelihood of the training graphs and their tree decompositions. 
Here we use Expectation-Maximization (EM)~\cite{dempster+al:1977}. In the E (Expectation) step, we use the Inside-Outside algorithm \cite{lari+young:1990} to compute the expected count of each subrule given the training data, and in the M (Maximization) step, we update the subrule probabilities by normalizing their expected counts.

\begin{figure}
\centering
\pgfdeclarepatternformonly{nort west lines}{\pgfqpoint{-0pt}{-0pt}}{\pgfqpoint{3pt}{3pt}}{\pgfqpoint{3pt}{3pt}}{
        \pgfsetlinewidth{0.4pt}
        \pgfpathmoveto{\pgfqpoint{0pt}{0pt}}
        \pgfpathlineto{\pgfqpoint{3pt}{3pt}}
        \pgfpathmoveto{\pgfqpoint{2.8pt}{-.2pt}}
        \pgfpathlineto{\pgfqpoint{3.2pt}{.2pt}}
        \pgfpathmoveto{\pgfqpoint{-.2pt}{2.8pt}}
        \pgfpathlineto{\pgfqpoint{.2pt}{3.2pt}}
        \pgfusepath{stroke}}

\pgfmathsetseed{83}

\begin{tikzpicture}
\fill[white, thick, draw=black,
      decoration={random steps,segment length=0.9cm,amplitude=0.6cm,pre=lineto,
      pre length=.5cm,post=lineto,post length=.4cm,},
      decorate,
      rounded corners=.3cm
    ] (0, 0) ellipse (1.6 and 1.6);
    
\pgfmathsetseed{13}
    
\fill[black!70!white, pattern=nort west lines, thick, draw=black,
      decoration={random steps,segment length=0.5cm,amplitude=0.3cm,pre=lineto,
      pre length=.2cm,post=lineto,post length=.1cm,},
      decorate,
      rounded corners=.2cm
    ] (0, 0) ellipse (1.25 and 1.25);    

\node[draw=black, circle, fill=black, scale=0.7] at (1,0.85) {};
\node[draw=black, circle, fill=black, scale=0.7] at (0,-1.3) {};
\node[draw=black, circle, fill=black, scale=0.7] at (-1.2,0.5) {};

\node[ fill=white, inner sep=0, opacity=.8, text opacity=1](x) at (-0.5,0.5) {$\eta,X_i$};
\framenode[3pt]{x}

\end{tikzpicture}
\pgfmathsetseed{83}
\begin{tikzpicture}
\fill[white, thick, draw=black,  pattern=nort west lines,
      decoration={random steps,segment length=0.9cm,amplitude=0.6cm,pre=lineto,
      pre length=.5cm,post=lineto,post length=.4cm,},
      decorate,
      rounded corners=.3cm
    ] (0, 0) ellipse (1.6 and 1.6);
    
\pgfmathsetseed{13}
    
\fill[white, thick, draw=black,
      decoration={random steps,segment length=0.5cm,amplitude=0.3cm,pre=lineto,
      pre length=.2cm,post=lineto,post length=.1cm,},
      decorate,
      rounded corners=.2cm
    ] (0, 0) ellipse (1.25 and 1.25);    

\node[draw=black, circle, fill=black, scale=0.7] at (1,0.85) {};
\node[draw=black, circle, fill=black, scale=0.7] at (0,-1.3) {};
\node[draw=black, circle, fill=black, scale=0.7] at (-1.2,0.5) {};

\node[ fill=white, circle, inner sep=0, opacity=.8, text opacity=1] at (-0.5,0.5) {$\eta,X_i$};

\end{tikzpicture}
\caption{Inside and outside probabilities. Here, a hyperedge labeled $X_i$ with three external nodes has generated a subgraph $\eta$. The inside probability (left) of $\eta,X_i$ is the probability of all the subderivations that generate subgraph $\eta$ from $X_i$. The outside probability (right) is the probability of all the partial derivations that generate the complement of subgraph $\eta$, with a hyperedge labeled $X_i$ in place of the subgraph.}
\label{fig:egg_ill}
\end{figure}
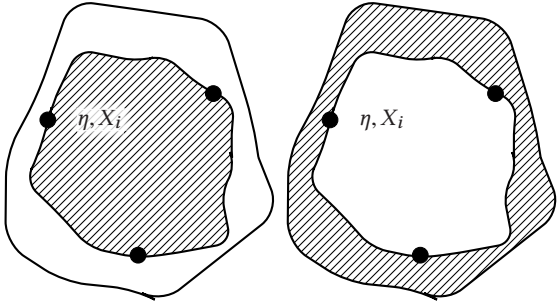

These expected counts can be computed efficiently using dynamic programming. 
Given a tree decomposition $T$, consider a bag $\eta$ and its corresponding subgraph $H_\eta$. The grammar extraction method of  Background Section~\ref{sec:extraction} assigns $H_\eta$ a nonterminal symbol, which we write $X$. Let $X_i$ be a subsymbol of $X$.
The \emph{inside} probability of $H_\eta$ with label $X_i$, written as $\inside{\eta}{X_i}$, is the total probability of all derivations starting from $X_i$ and ending in $H_\eta$. The \emph{outside} probability of $H_\eta$ with label $X_i$, written as $\outside{\eta}{X_i}$, is the total probability of all derivations starting from $S$ and ending in $H$ with $H_\eta$ replaced with a hyperedge labeled $X_i$. See Figure~\ref{fig:egg_ill}.

\begin{figure}
\centering
\pgfmathsetseed{83}

\begin{tikzpicture}
\fill[white, thick, draw=black,
      decoration={random steps,segment length=0.9cm,amplitude=0.6cm,pre=lineto,
      pre length=.5cm,post=lineto,post length=.4cm,},
      decorate,
      rounded corners=.2cm
    ] (0, 0) ellipse (1.6 and 1.6);
    
\pgfmathsetseed{13}
    
\fill[black!70!white, pattern=nort west lines, thick, draw=black,
      decoration={random steps,segment length=0.5cm,amplitude=0.3cm,pre=lineto,
      pre length=.2cm,post=lineto,post length=.1cm,},
      decorate,
      rounded corners=.1cm
    ] (0, 0) ellipse (1.25 and 1.25);    

\node[draw=black, circle, fill=black, scale=0.7] at (1,0.85) {};
\node[draw=black, circle, fill=black, scale=0.7] at (0,-1.3) {};
\node[draw=black, circle, fill=black, scale=0.7] at (-1.2,0.5) {};

\node[ fill=white, inner sep=0, opacity=.8, text opacity=1](x) at (-0.5,0.9) {$\eta,X_i$};
\framenode[3pt]{x}

\pgfmathsetseed{12}

\node[ fill=white, inner sep=0, opacity=.8, text opacity=1](x) at (-0.6,0.0) {$\eta_1,Y_j$};
\framenode[3pt]{x}

\fill[fill opacity=0, thick, draw=black,
      decoration={random steps,segment length=0.3cm,amplitude=0.2cm,pre=lineto,
      pre length=.1cm,post=lineto,post length=.1cm,},
      decorate,
      rounded corners=.2cm
    ] (-0.6, 0) ellipse (0.5 and 0.4);    

\node[draw=black, circle, fill=black, scale=0.6] at (-0.5,-0.55) {};
\node[draw=black, circle, fill=black, scale=0.6] at (-0.35,0.3) {};

\node[ fill=white, inner sep=0, opacity=.8, text opacity=1](x) at (0.6,0.0){$\eta_2,Z_k$};
\framenode[3pt]{x}

\pgfmathsetseed{2}

\fill[fill opacity=0, thick, draw=black,
      decoration={random steps,segment length=0.3cm,amplitude=0.2cm,pre=lineto,
      pre length=.1cm,post=lineto,post length=.1cm,},
      decorate,
      rounded corners=.2cm
    ] (0.6, 0) ellipse (0.4 and 0.4);    

\node[draw=black, circle, fill=black, scale=0.6] at (0.9,-0.45) {};
\node[draw=black, circle, fill=black, scale=0.6] at (1.0,0.3) {};
\node[draw=black, circle, fill=black, scale=0.6] at (0.10,0.1) {};

\end{tikzpicture}
\begin{tikzpicture}
\pgfmathsetseed{83}
\fill[pattern=nort west lines, thick, draw=black,
      decoration={random steps,segment length=0.9cm,amplitude=0.6cm,pre=lineto,
      pre length=.5cm,post=lineto,post length=.4cm,},
      decorate,
      rounded corners=.3cm
    ] (0, 0) ellipse (1.6 and 1.6);
    
\pgfmathsetseed{13}
    
\fill[fill opacity=0, thick, draw=black,
      decoration={random steps,segment length=0.5cm,amplitude=0.3cm,pre=lineto,
      pre length=.2cm,post=lineto,post length=.1cm,},
      decorate,
      rounded corners=.2cm
    ] (0, 0) ellipse (1.25 and 1.25);    

\node[draw=black, circle, fill=black, scale=0.7] at (1,0.85) {};
\node[draw=black, circle, fill=black, scale=0.7] at (0,-1.3) {};
\node[draw=black, circle, fill=black, scale=0.7] at (-1.2,0.5) {};

\node[fill=white, inner sep=0, opacity=.8, text opacity=1](x) at (-0.5,0.9) {$\eta,X_i$};
\framenode[3pt]{x}

\pgfmathsetseed{12}

\fill[white, thick, draw=black,
      decoration={random steps,segment length=0.3cm,amplitude=0.2cm,pre=lineto,
      pre length=.1cm,post=lineto,post length=.1cm,},
      decorate,
      rounded corners=.2cm
    ] (-0.6, 0) ellipse (0.5 and 0.4);    

\node[draw=black, circle, fill=black, scale=0.6] at (-0.5,-0.55) {};
\node[draw=black, circle, fill=black, scale=0.6] at (-0.35,0.3) {};
    
\node[  inner sep=0, text opacity=1] (x)  at (-0.6,0.0) {$\eta_1,Y_j$};

\node[ fill=white, inner sep=0, opacity=.8, text opacity=1](x) at (0.6,0.0) {$\eta_2,Z_k$};
\framenode[3pt]{x}

\pgfmathsetseed{2}

\fill[fill opacity=0, thick, draw=black,
      decoration={random steps,segment length=0.3cm,amplitude=0.2cm,pre=lineto,
      pre length=.1cm,post=lineto,post length=.1cm,},
      decorate,
      rounded corners=.2cm
    ] (0.6, 0) ellipse (0.4 and 0.4);    

\node[draw=black, circle, fill=black, scale=0.6] at (0.9,-0.45) {};
\node[draw=black, circle, fill=black, scale=0.6] at (1.0,0.3) {};
\node[draw=black, circle, fill=black, scale=0.6] at (0.10,0.1) {};

\end{tikzpicture}
\caption{Computation of inside and outside probabilities. Here, a hyperedge labeled $X_i$ has been rewritten with a rule rhs with two hyperedges labeled $Y_j$ and $Z_k$. At left, the inside probability of $\eta,X_i$ is incremented by the product of the rule and the inside probabilities of $\eta_1,Y_j$ and $\eta_2,Z_k$. At right, the outside probability of $\eta_1,Y_j$ is incremented by the product of the outside probability of $\eta,X_i$, the rule, and the inside probability of $\eta_2,Z_k$.}
\label{fig:egg_calc}
\end{figure}
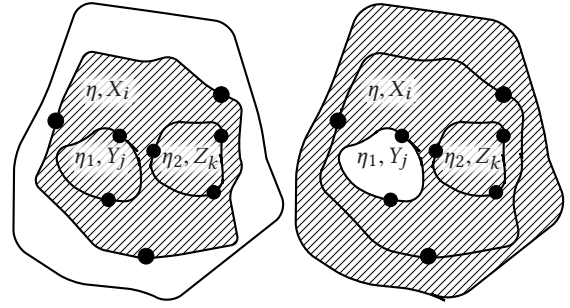

The inside probabilities can be calculated recursively, from smaller subgraphs to larger subgraphs. 
We assume that bag $\eta$ has at most two children, which follows if $T$ is in Chomsky Normal Form~\cite{chomsky1959certain}. If $\eta$ has two children, let $\eta_1$ and $\eta_2$ be the children, let $Y$ and $Z$ be the labels of $H_{\eta_1}$ and $H_{\eta_2}$, and let $Y_j$ be and $Z_k$ be subsymbols of $Y$ and $Z$. Then the inside probability of $H_\eta$ with subsymbol $X_i$ is defined by:
\begin{align*}
    \inside{\eta}{X_i} &= \sum_{j, k} \ruleprob{X_i\rightarrow Y_j Z_k} \, \inside{\eta_1}{Y_j} \, \inside{\eta_2}{Z_k} \\
\intertext{and similarly if $\eta$ has only one child:}    
\inside{\eta}{X_i} &= \sum_{j} \ruleprob{X_i\rightarrow Y_j} \, \inside{\eta_1}{Y_j} \\
\intertext{or no children:}
\inside{\eta}{X_i} &= \ruleprob{X_i\rightarrow \epsilon}.
\end{align*}    

The outside probabilities are calculated top-down. 
If a bag~$\eta$ has two children, then the outside probabilities of its children are defined by:
\begin{align*}    
    \outside{\eta_1}{Y_j} &= \sum_{i, k} \ruleprob{X_i\rightarrow Y_j Z_k} \, \outside{\eta}{X_i} \, \inside{\eta_2}{Z_k} \\
    \outside{\eta_2}{Z_k} &= \sum_{i, j} \ruleprob{X_i\rightarrow Y_j Z_k} \, \outside{\eta}{X_i} \, \inside{\eta_1}{Y_k}. \\
\intertext{See Figure~\ref{fig:egg_calc} for an illustration of this computation. Similarly, if $\eta$ has only one child:}
    \outside{\eta_1}{Y_j} &= \sum_{i} \ruleprob{X_i\rightarrow Y_j} \, \outside{\eta}{X_i}.
\end{align*}

In the Expectation step, we compute the posterior probability of each subrule at each bag of each training tree decomposition~$T$:
\begin{align*}
    P(\eta, X_i\rightarrow Y_j  Z_k\mid T) &= \frac1{P(T)}\outside{\eta}{X_i}  \ruleprob{X_i\rightarrow Y_j Z_k} \cdot{} \\ &\qquad \inside{\eta_1}{Y_j}  \inside{\eta_2}{Z_k} \\
    P(\eta, X_i \rightarrow Y_j \mid T) &= \frac1{P(T)}\outside{\eta}{X_i}  \ruleprob{X_i \rightarrow Y_j}  \inside{\eta_1}{Y_j} \\
    P(\eta, X_i \rightarrow \epsilon \mid T) &= \frac1{P(T)}\outside{\eta}{X_i}  \ruleprob{X_i \rightarrow \epsilon}
\end{align*}
where $P(T) = \inside{\eta_0}{S}$ and $\eta_0$ is the root bag of $T$.
The expected count of each subrule is calculated by summing over the posterior probability of the rule over all nodes of all training trees:
\begin{equation*}
    E[c(X_i\rightarrow \alpha)] = \sum_{\text{trees $T$}} \sum_{\eta \in T} P(\eta, X_i\rightarrow \alpha\mid T) 
\end{equation*}
where $\alpha$ is any right-hand side.

In the Maximization step, we use the expected counts calculated above to update the rule probabilities:
\begin{align*}
    \ruleprob{X_i\rightarrow \alpha} &:= \frac{E[c(X_i\rightarrow \alpha)]}{\sum_{\alpha'} E[c(X_i \rightarrow \alpha')]}.
\end{align*}

These probabilities are then used to repeat the E step. The method is guaranteed to converge to a local maximum of the likelihood function, but not necessarily to a global maximum.

\section{Evaluation}

Current research in graph modelling and graph generation evaluate their results by comparing the generated graphs with the original graph by aggregate properties like degree distribution, clustering coefficients, or diameter~\cite{aguinaga2016growing,leskovec2010kronecker,aguinaga2016infinity,seshadhri2012community,leskovec2005graphs,chakrabarti2006graph}. There are two potential problems with such metrics. First, these metrics do not test how well the model generalizes to model other graphs that represent similar data. Second, they are heuristics from which a generated graph's ``goodness'' is difficult to define or standardize. We discuss and address both of these problems in this section.

\subsection{Train/Test Data}

Comparing generated graphs with the original graph cannot test how well the model generalizes to other graphs that represent similar data or different versions of the same network phenomena. To see why, consider the extreme case, in which a model simply memorizes the entire original graph. Then, the generated graphs are all identical to the original graph and therefore score perfectly according to these metrics. This is akin to overfitting a machine learning classifier on training data and then testing it on the same data, which would not reveal whether the model is able to generalize to unseen instances. 

In standard data mining and machine learning tasks, the overfitting problem is typically addressed through cross-validation or by evaluating on heldout test data sets. In the present work, we adapt the idea of using heldout test data to evaluate graph grammars. 
In experiments on synthetic graphs, this means that we generate two random graphs using the same model and parameters; we designate one as the training graph and the other as the test graph. In experiments on real world graphs, we identify two graphs that represent the same phenomenon, \eg, citations or collaborations, and we mark one as the training graph and one as the test graph. 

In reality, we might not be able to find test graphs that have similar properties as the training graph. Fortunately, cross-validation can also be adapted to cases where no test graph is available by using disjoint subgraph samples from a single graph.

\subsection{Likelihood}

In addition to the possibility of overfitting, high-level aggregations of graph properties may not always be good comparators of two or more graphs. Indeed, examples abound in related literature showing how vastly different graphs can share similar aggregate statistics~\cite{yaveroglu2014revealing}. We propose, as an additional metric, to evaluate models by using them to measure the likelihood of a test graph or graphs. Intuitively, this measures how well a model extracted from the training graph generalizes to a test graph. If the model simply memorizes the entire training graph, then it will have zero likelihood (the worst possible) on the test graph. If the model is better able to generalize to new graphs, then it will have higher likelihood on the test graph.

Unfortunately, it is not always computationally feasible to compute the likelihood of graphs under previous models. But with HRGs, it can be computed in linear time given a tree decomposition. (It would also be possible, but slower, to sum the probabilities of \emph{all} possible tree decompositions \cite{chiang2013parsing}.)  The likelihood on a test graph is simply $\inside{\eta_0}{S}$, where $\eta_0$ is the root of the tree decomposition. Note that the model probabilities are estimated from the training graphs, even when computing likelihood on test graphs. As this number is usually very small, it's common to take logs and deal with log-likelihoods.

\subsection{Smoothing}
A problem arises, however, if the test graph uses a rule that does not exist in the grammar extracted from the training graph. Then the inside probability will be zero (or a log-probability of $-\infty$). This is because an HRG missing any necessary rules to construct the test graph cannot generate the test graph exactly, and therefore results in a zero probability.

In this case, we would still like to perform meaningful comparisons between models, if possible. So we apply smoothing as follows. To test an HRG $H$ on a test graph, we first extract an HRG, $H^\prime$, from the test graph using the latent-variable HRG method. Define an \emph{unknown rule} to be a rule in $H^\prime$ but not in $H$. Then for each unknown rule, we add the rule to $H$ with a probability of $\epsilon$. We can then compute the log likelihood on the test graph under the augmented grammar $H \cup H^\prime$. The final test log likelihood is calculated as
\begin{align*}
    L = L_{H \cup H^\prime} - c(H^\prime \setminus H) \cdot \log \epsilon
\end{align*}
where $L_{H \cup H^\prime}$ is the log likelihood of the test graph under the augmented grammar, $c(H^\prime \setminus H)$ is the number of times that unknown rules are used in the test graph, and $\epsilon$ is the probability of each unknown rule. Note that as long as $\epsilon$ is much smaller than the probability of any known rule, its value is irrelevant because $L$ does not depend on it.

Ideally, we would like the number of unknown rules to be zero. In our experiments, we find that increasing the number of training graphs and/or decreasing the size of the training graphs can bring this number to zero or close to zero. Note that if two HRGs have differing sets of unknown rules, then it is \emph{not} meaningful to compare their log-likelihood on test graphs. But if two HRGs have identical sets of unknown rules, then their log-likelihoods can be meaningfully compared. We will exploit this fact when evaluating models with latent variables in the next section.

\section{Experiments}

In this section we test the ability of the latent-variable HRG (laHRG) to generate graphs using the train-test framework described above. We vary $n$, the number of subsymbols that each nonterminal symbol is split into, from 1 to 4. Note that the 1-split laHRG model is identical to the original HRG method. By varying the number of splits, we will be able to find the value that optimizes the test likelihood. We have provided all of the source code and data analysis scripts at \url{https://github.com/cindyxinyiwang/laHRG}.

\subsection{Setup}
Given a training graph, we extract and train a latent-variable HRG from the graph. Then we evaluate the goodness of the grammar by calculating the log likelihood that the test graph could be generated from the grammar. 

\begin{table}
\tabcolsep=1.5pt\relax
 \small
\caption{Datasets used in experiments}
\label{table: real-world-graphs}
\centering
\begin{tabular}{lllrr} 
\toprule
 & Type & Name &  Nodes & Edges \\ 
\midrule
 \multirow{4}{*}{Synth} &
 \multirow{2}{*}{Barabasi-Albert}
 & train-ba &  30,000 & 59,996 \\ 
 & & test-ba &  30,000 & 59,996 \\
 \cmidrule{2-5}
 & \multirow{2}{*}{Watts-Strogatz}
 & train-ws & 30,000 & 60,000 \\
 & & test-ws  & 30,000 & 60,000 \\

\midrule

 \multirow{10}{*}{Real} &
 \multirow{2}{*}{Citation}
 & train-cit-HepTh &  27,770 & 352,807 \\ 
 & & test-cit-HepPh &  34,546 & 421,578 \\
 \cmidrule{2-5}
 & \multirow{2}{*}{Internet}
 & train-as-topo & 34,761 & 171,403 \\
 & & test-as-733  & 6,474 & 13,895 \\
 \cmidrule{2-5}
 & \multirow{2}{*}{Purchase}
 & train-amazon0312  & 400,727 & 3,200,440 \\
 && test-amazon0302 & 262,111 & 1,234,877 \\ 
 \cmidrule{2-5}
 & \multirow{2}{*}{Wikipedia}
 & train-wiki-vote  & 7,115 & 103,689 \\ 
 & & test-wiki-talk & 2,394,385 & 5,021,410 \\ 
\bottomrule
\end{tabular}
\end{table}

For evaluation, we need to make sure that the test graph is disjoint from the training graph to guard against overfitting. Here we introduce two techniques to achieve that: 1. we partition a single graph into two disjoint parts so that they do not have overlapping vertices. Then we train laHRG from one part of the graph, and calculate the log likelihood of the other disjoint part; 2. we choose 2 graphs of the same type, and use one for training and the other for evaluation.

We evaluate laHRG with log likelihood metric for both evaluation methods mentioned above on 6 types of graphs: 2 synthetic graphs (generated from random graph generators), and 4 real world graphs. For the first evaluation method, we use 1 graph, each from 6 different types of graphs, and partition the graph for training and testing purpose. For the second evaluation method, we do not partition the training graph, but choose 1 additional graph from each type of graphs for testing. The graphs were obtained from the SNAP\footnote{\url{https://snap.stanford.edu/data}} and KONECT\footnote{\url{http://konect.uni-koblenz.de}} graph repositories and are listed in Table~\ref{table: real-world-graphs}. 

Many of these graphs are too large for a tree decomposition to be calculated. Instead, we randomly sampled a set of fixed-size subgraphs from the training graph and a set of fixed-size subgraphs from the test graph. Besides the concern from the calculation of large tree decompositions, sampling multiple graphs is also important for the extraction of a broad set of rules. Recall that if a single rule required to generate the test graph is not found within the training graph, then the likelihood will be 0. Therefore, large test graphs would require many more training graphs in order to reduce (or hopefully eliminate) the need for smoothing.

In all experiments, we extract 500 samples of size-25 subgraphs from the training graph. We extract an HRG from each size-25 subgraph, perform nonterminal splitting and EM training. The 500 HRGs are then combined and their weights are normalized to create the final laHRG model. We also take 4 samples of size-25 subgraphs from the test graph, calculate the log-likelihood of each under the laHRG model, and report the mean log-likelihood and confidence interval. 

We chose these parameters empirically such that there is no need for smoothing. 
If we were to increase the subgraph size for the test graphs, then we would also need to increase the number of training graph samples or rely on smoothing to ensure non-zero likelihood. 

To compute tree decompositions, we used a reimplementation of the QuickBB algorithm \cite{gogate-dechter-uai04}, with only the ``simplicial'' and ``almost-simplicial'' heuristics. 

\subsection{Log-Likelihood Results} \label{sec:loglikelihood}

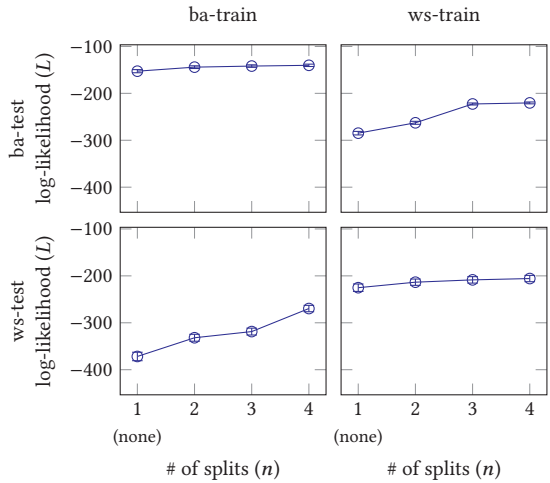
\begin{figure}[t]
\centering
\begin{tikzpicture} \small
 \begin{groupplot}[compat=1.5,
    group style = {group size = 2 by 2,
        horizontal sep=0.2cm,
        vertical sep=0.2cm,
        xticklabels at=edge bottom,
        yticklabels at=edge left,
        },
    width = 1.7in, height = 1.5in,
    enlarge y limits=0.01,
    ymin = -450,
    ymax = -100,
    xlabel near ticks,
    ylabel style={align=center},
    xtick ={1,2,3,4}, 
xticklabel style={align=center}, xticklabels={{1\\\footnotesize (none)},2,3,4},    ytick={-100, -200, -300, -400, -500},
    ]
    \nextgroupplot[title={ba-train}, ylabel={ba-test \\ log-likelihood ($L$)}]
    
    \addplot[blue, mark=o, legend style={draw=none}, error bars/.cd, y dir=both, y explicit,] coordinates {
    
    (1,	-152.776019584) +=(0, 3.348912902) -=(0,3.348912902)
    (2,	-144.175275196) +=(0, 2.89493062) -=(0,2.89493062)
    (3,	-141.885491821) +=(0, 2.697075707) -=(0,2.697075707)
    (4,	-140.259644003) +=(0, 2.528982413) -=(0,2.528982413)
    };

    \nextgroupplot[title={ws-train}    ]
    
    \addplot[blue, mark=o, legend style={draw=none}, error bars/.cd, y dir=both, y explicit,] coordinates {
    
    (1,	-284.872592824) +=(0, 3.764946522) -=(0,3.764946522)
    (2,	-262.7882677) +=(0, 3.165217459) -=(0,3.165217459)
    (3,	-222.670338731) +=(0, 2.146449348) -=(0,2.146449348)
    (4,	-220.386074332) +=(0, 1.992801489) -=(0,1.992801489)
    };
    
    \nextgroupplot[xlabel = {\# of splits ($n$)}, ylabel={ws-test \\ log-likelihood ($L$)}]

    \addplot[blue, mark=o, legend style={draw=none}, error bars/.cd, y dir=both, y explicit] coordinates {
    (1, -371.776420634) +=(0, 9.022746443) -=(0, 9.022746443)
    (2, -331.900946475) +=(0, 7.387076869) -=(0, 7.387076869)
    (3, -318.732979991) +=(0, 7.318296779) -=(0, 7.318296779)
    (4, -269.641295246) +=(0, 6.815057562) -=(0, 6.815057562)
    };
    
    \nextgroupplot[      xlabel = {\# of splits ($n$)},
  ]

    \addplot[blue, mark=o, legend style={draw=none}, error bars/.cd, y dir=both, y explicit] coordinates {
    (1, -225.210256037) +=(0, 7.94523990234) -=(0, 7.94523990234)
    (2, -213.470608835) +=(0, 6.72806504513) -=(0, 6.72806504513)
    (3, -208.533064422) +=(0, 6.89039736158) -=(0, 6.89039736158)
    (4, -205.710844636) +=(0, 6.73518368397) -=(0, 6.73518368397)
    };
\end{groupplot}
\end{tikzpicture}
\caption{On synthetic graphs, splitting nonterminal symbols ($n \geq 2$) always improves log-likelihood on the test graph, as compared to no splitting ($n=1$). We did not observe overfitting up to $n=4$. Left/right column: train on Barabasi-Albert (ba) or Watts-Strogatz (ws). Top/bottom row: test on ba/ws. 
}
\label{fig:gen_loglikelihood}
\end{figure}

This section explains the performance of laHRG in terms of log-likelihood metric on test graph for two different train-test split methods mentioned in the previous section. We mainly analyze the results for the second method: train on one graph and test on another graph of the same type. The first method has similar results, and we include them here to show that our evaluation method also works for graphs that are difficult to find different test graphs of the same type.

\subsubsection{Validate on Different Graph of Same Type}
We first show the log-likelihood results on synthetic datasets. The two random graph models, the Barabasi-Albert graph and Watts-Strogatz graph, generate very different graph types. 
The four panels in Fig.~\ref{fig:gen_loglikelihood} show the log-likelihood results of four combinations of training graphs and test graphs. Higher is better. 

As a sanity check, we also trained an laHRG model on a Barabasi-Albert graph and tested it against a Watts-Strogatz graph and vice versa. We expect to see much lower log-liklihood scores because the laHRG trained on one type of graph should be different than another type of graph. The top-right and bottom-left panels in Fig.~\ref{fig:gen_loglikelihood} show that this is indeed the case; the log-likelihood measure and the laHRG model pass the sanity check.

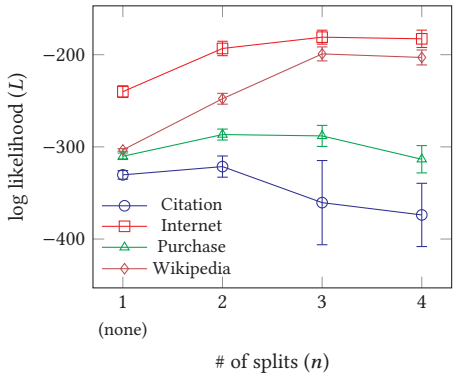
\begin{figure}
\centering

\begin{tikzpicture} 
\small
\begin{axis}[
    width = 2.5 in, height = 2.1in,
    enlarge y limits=0.01,
    ylabel = {log likelihood ($L$)},
    xlabel = {\# of splits ($n$)},
    ymax=-150,
    ymin=-450,
    ytick={0, -100, -200, -300, -400},
    xtick ={1,2,3,4},
    xticklabel style={align=center}, xticklabels={{1\\\footnotesize (none)},2,3,4},
    label style={},
    legend style={draw=none, fill=none, anchor=south west, at={(0,0)},font={\footnotesize},row sep=-2pt},
    ]
\addplot[blue, mark=o, legend style={draw=none}, error bars/.cd, y dir=both, y explicit,] coordinates {

(1,	-330.377882455) +=(0, 4.69050147) -=(0,4.69050147)
(2,	-321.396849871) +=(0, 11.494855669) -=(0,11.494855669)
(3,	-360.512034633) +=(0, 45.673480302) -=(0,45.673480302)
(4,	-373.843602984) +=(0, 34.314993466) -=(0,34.314993466)

};
\addlegendentry{Citation}

\addplot[red, mark=square, legend style={draw=none}, error bars/.cd, y dir=both, y explicit] coordinates {
(1, -240.089110038) +=(0, 6.344192634) -=(0, 6.344192634)
(2, -193.238534265) +=(0, 7.821099197) -=(0, 7.821099197)
(3, -180.990169402) +=(0, 7.56719803) -=(0, 7.56719803)
(4, -182.766147277) +=(0, 9.419074824) -=(0, 9.419074824)
};
\addlegendentry{Internet}

\addplot[green, mark=triangle, legend style={draw=none}, error bars/.cd, y dir=both, y explicit,] coordinates {

(1,	-310.203075987) +=(0, 3.52226633523) -=(0,3.52226633523)
(2,	-286.641433419) +=(0, 5.92901739522) -=(0,5.92901739522)
(3,	-288.116076275) +=(0, 11.4118894697) -=(0,11.4118894697)
(4,	-313.437543514) +=(0, 14.7993979945) -=(0,14.7993979945)

};
\addlegendentry{Purchase}

\addplot[purple, mark=diamond, legend style={draw=none}, error bars/.cd, y dir=both, y explicit] coordinates {
(1, -303.32551683) +=(0, 1.41367759655) -=(0, 1.41367759655)
(2, -247.799221309) +=(0, 5.88138155723) -=(0, 5.88138155723)
(3, -199.156949108) +=(0, 7.59099397631) -=(0, 7.59099397631)
(4, -202.987068704) +=(0, 8.11454742731) -=(0, 8.11454742731)
};
\addlegendentry{Wikipedia}

\end{axis}
\end{tikzpicture}
\caption{On real-world graphs, splitting nonterminal symbols ($n \geq 2$) always improves log-likelihood on the test graph, as compared to no splitting ($n=1$), peaking at 2 or 3 splits and then dropping due to overfitting. Error bars indicate 95\% confidence intervals. Higher is better.}
\label{fig:real_loglikelihood}
\end{figure}

Next we extracted and tested the laHRG model on real world graphs. The log-likelihood results are illustrated in Fig.~\ref{fig:real_loglikelihood} for laHRG models of up to 4-splits. 
We find that the log-likelihood scores peak at $n$ = 2 or 3 and then decreases when $n$ = 4.  

Recall that laHRG is the same as HRG~\cite{aguinaga2016growing} when $n$ = 1. Based on the results from Fig.~\ref{fig:real_loglikelihood}, we find that splitting does indeed increase HRG's ability to generate the test graph. However, as increasing $n$ shows diminishing returns and sometimes decreases performance. The decrease in log likelihood when $n>2$ is caused by model overfitting. The increase in node-splitting allows laHRG to fine-tune the rule probabilities to the training graph, which we find does not always generalize to the test graph.

\subsubsection{Validate on Disjoint Subgraph}

\begin{figure}
\centering
\begin{tikzpicture} \small
 \begin{groupplot}[compat=1.5,
    group style = {group size = 2 by 3,
        horizontal sep=0.2cm,
        vertical sep=0.8cm,
        xticklabels at=edge bottom,
        yticklabels at=edge left,
        xlabels at=edge bottom,
        ylabels at=edge left,
        },
    width = 1.5in, height = 1.3in,
    title style={yshift=-0.2cm,font={\strut}},
    ylabel = {log likelihood ($L$)},
    xlabel = {\# of splits ($n$)},
    ymax=-130,
    ymin=-450,
    ytick={0, -100, -200, -300, -400},
    xtick ={1,2,3,4},
    legend style={draw=none, fill=none},
]

\nextgroupplot[title={Barabasi-Albert}]
\addplot[blue, mark=o, error bars/.cd, y dir=both, y explicit,] coordinates {
(1,	-155.631143309) +-=(0, 3.82698773388) 
(2,	-151.722372185) +-=(0, 3.52264365898) 
(3,	-146.886502229) +-=(0, 2.89441681257) 
(4,	-147.850922997) +-=(0,2.9916841705) 
};

\nextgroupplot[title={Watts-Strogatz}]    
\addplot[blue, mark=o, error bars/.cd, y dir=both, y explicit,] coordinates {
(1,	-222.425187963) +-=(0,1.53411789885) 
(2,	-210.59409728) +-=(0, 1.34474414626) 
(3,	-202.93637126) +-=(0, 1.18361012707) 
(4,	-200.832165137) +-=(0, 1.17028811283) 
};

\nextgroupplot[title={Citation}]
\addplot[blue, mark=o, error bars/.cd, y dir=both, y explicit,] coordinates {
(1,	-321.790955961) +-=(0, 5.2294202336)
(2,	-317.403700532) +-=(0, 19.9871124624)
(3,	 -328.724512624) +-=(0, 30.9839803867)
(4,	-367.774216258) +-=(0, 57.2217981701)
};

\nextgroupplot[title={Internet}]
\addplot[blue, mark=o, error bars/.cd, y dir=both, y explicit,] coordinates {
(1,	-311.277451323) +-=(0, 7.58850446644) 
(2,	-295.615429857) +-=(0, 18.9376543073)
(3,	-286.333999359) +-=(0, 41.8512539186) 
(4,	-286.530959369) +-=(0, 28.3251808711)
};

\nextgroupplot[title={Purchase}]    
\addplot[blue, mark=o, error bars/.cd, y dir=both, y explicit,] coordinates {
(1,	-311.277451323) +-=(0, 7.58850446644) 
(2,	-295.615429857) +-=(0, 18.9376543073)
(3,	-286.333999359) +-=(0, 41.8512539186) 
(4,	-286.530959369) +-=(0, 28.3251808711) 
};

\nextgroupplot[title={Wikipedia}]
\addplot[blue, mark=o, error bars/.cd, y dir=both, y explicit,] coordinates {
(1, -330.35599518) +-=(0, 8.19357421038) 
(2,  -314.702775217) +-=(0, 32.168730057) 
(3, -300.590282921) +-=(0, 41.5141941239)
(4,  -359.971176298) +-=(0, 86.8555122164)
};

\end{groupplot}

\end{tikzpicture}
\caption{Loglikelihood on subgraphs that are disjoint from training graphs. Trends similar to Fig.~\ref{fig:real_loglikelihood} and Fig.~\ref{fig:gen_loglikelihood} can be observed with this method. Error bars indicate 95\% confidence intervals. Higher is better.}
\label{fig:single_loglikelihood}
\end{figure}
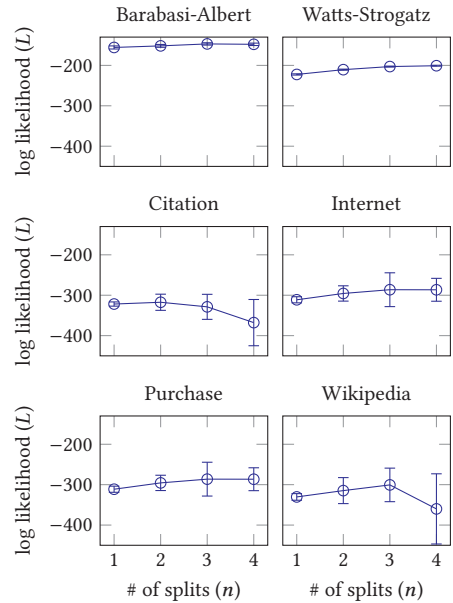

If it is difficult to find a test graph of the same type with the training graph, it is still possible to evaluate laHRG with the log likelihood metric. We can partition the graph into two disjoint parts, and use one for training and the other for testing. Fig.~\ref{fig:single_loglikelihood} shows the log likelihood of the test subgraphs that are disjoint from the training graphs. Again, splitting nonterminal symbols increase the log likelihood on the test graph, but as the number of splits ($n$) increases, log likelihood decreases due to overfitting.

\subsection{Comparing against other Graph Generators}

The log-likelihood metric is a principled approach to calculating the performance of a graph generator. Unfortunately, other graph generators are not capable of performing this type of likelihood calculation. In order to compare the laHRG graph model to other state of the art graph generators including the Kronecker~\cite{leskovec2010kronecker} and Chung-Lu~\cite{chung+lu} models, we revert to traditional graph metrics to compare a generated graph to a test graph(not against the original graph). 


Among the many choices for heuristic graph comparison metrics, we chose the degree distribution and graphlet correlation distance (GCD).

Recall that the sampling of 25-node subgraphs was necessary to ensure a non-zero probability for the log likelihood evaluation. No such requirement exists for evaluation on GCD. 
Nevertheless, to maintain an apples to apples comparison, we performed similar graph sampling methods for degree distribution distance and GCD: we trained Kronecker and Chung-Lu models on a 25-node subgraph from the training graph, generated a 25-node graph, compared the generated graph against a 25-node subgraph of the test graph, and repeated this process 500 times. 

As a baseline, we also compared the training and test graph directly to get a basic sense of their similarity. So, we directly compared 25-node subgraphs from the training graph to 25-node subgraphs of the test graph without any model. We repeated this direct comparison 500 times and report the mean and 95\% confidence interval. 
We call this the ``Direct'' comparison because it does not involve any graph generation. 

\subsubsection{Degree Distribution Distance}

In the present work we apply the degree distribution distance of Pr{\v{z}}ulj~\cite{prvzulj2007biological} to compare two or more degree distributions. Lower degree distribution distance between two graphs means they are more similar. 

which is defined as follows. Given a graph $H$, we first scale and normalize the degree distribution of $H$:

$$S_H(k) = \frac{d_H(k)}{k}$$
$$T_H = \sum_{k=1}^\infty{S_H(k)}$$
$$N_H(k) = \frac{S_H(k)}{T_H}$$

\noindent in order to reduce the effect of higher degree nodes, where $d_H(K)$ is the number of nodes in $H$ that have a degree of $k$. Then we calculate the distance between two degree distributions $D\left(d_H, d_{H^\prime}\right)$ as:

$$D\left(d_H, d_{H^\prime}\right) = \frac{1}{\sqrt{2}}\sqrt{ \sum^\infty_{k=1}\left( N_H(k) - N_{H^\prime}(k) \right)^2 },$$

\noindent which is essentially a normalized sum of squares between the two distributions. We call this metric the degree distribution distance. Because this is a ``distance'' metric low values indicate high similarity.

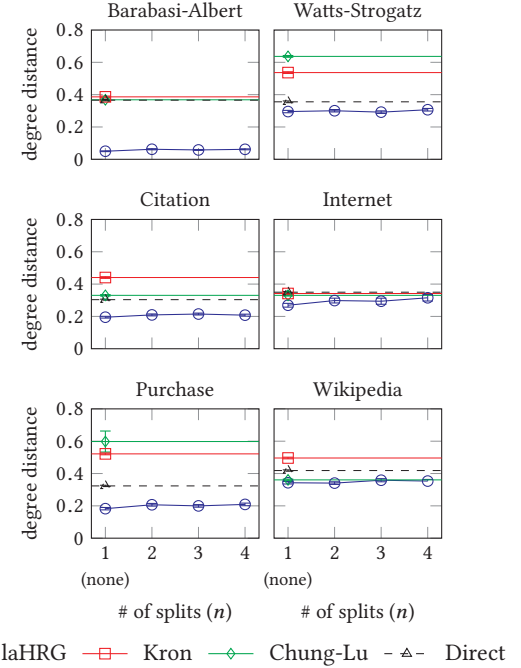
\begin{figure}[t]
\centering
\begin{tikzpicture} \small

\begin{groupplot}[compat=1.5,
    group style={group size=2 by 3,        
                 horizontal sep=0.2cm,
                 vertical sep=0.8cm,
                 xticklabels at=edge bottom,
                 yticklabels at=edge left,
                 xlabels at=edge bottom,
                 ylabels at=edge left,
    },
    ymax = 0.8,
    ymin = 0.0,
    width=1.5in,height=1.3in,
    xlabel={\# of splits ($n$)}, xtick={1,2,3,4}, 
    xticklabel style={align=center}, xticklabels={{1\\\footnotesize (none)},2,3,4},
    ylabel={degree distance},
    legend style={draw=none, fill=none},
    title style={yshift=-2ex,font={\strut}}]

\nextgroupplot[title={Barabasi-Albert}]
\addplot[color=blue, mark=o, error bars/.cd, y dir=both, y explicit,] coordinates {
(1,	0.0499059087944) +-=(0, 0.0023452484588) 
(2,	0.0621711660577) +-=(0, 0.00339441762444) 
(3,	0.05797243578541) +-=(0, 0.00271622106938) 
(4,	0.0617850613676) +-=(0,0.00284987473066) 
};

\addplot[color=green, mark=diamond, error bars/.cd, y dir=both, y explicit] coordinates { 
(1, 0.36822379)	+-=(0, 	0.004995107)
};
\draw[color=green, ultra thin] (axis cs:\pgfkeysvalueof{/pgfplots/xmin},0.36822379)--(axis cs:\pgfkeysvalueof{/pgfplots/xmax},0.36822379);

\addplot[color=red, mark=square, error bars/.cd, y dir=both, y explicit,] coordinates { 
(1, 0.386361882)	+-=(0,	0.004955996)
};

\draw[color=red, ultra thin] (axis cs:\pgfkeysvalueof{/pgfplots/xmin},0.386361882)--(axis cs:\pgfkeysvalueof{/pgfplots/xmax},0.386361882);

\addplot[color=black, dashed, mark=triangle, error bars/.cd, y dir=both, y explicit,] coordinates { 
(1, 0.365828538)	+-=(0,	0.0041313959)
};
\draw[color=black, dashed, ultra thin] (axis cs:\pgfkeysvalueof{/pgfplots/xmin},0.365828538)--(axis cs:\pgfkeysvalueof{/pgfplots/xmax},0.365828538);

\nextgroupplot[title={Watts-Strogatz}]
\addplot[color=blue, mark=o, error bars/.cd, y dir=both, y explicit,] coordinates {
(1,	0.29535562192) +-=(0, 0.00737649250746) 
(2,	0.30045737331) +-=(0, 0.00746358458345) 
(3,	 0.291778527817) +-=(0, 0.007951449678785) 
(4, 0.306640009156) +-=(0,0.00801682385777) 
};    

\addplot[color=green, mark=diamond, error bars/.cd, y dir=both, y explicit,] coordinates { 
(1, 0.63674768)	+-=(0,	0.00547354)
};
\draw[color=green, ultra thin] (axis cs:\pgfkeysvalueof{/pgfplots/xmin},0.63674768)--(axis cs:\pgfkeysvalueof{/pgfplots/xmax},0.63674768);
\addplot[color=red, mark=square, error bars/.cd, y dir=both, y explicit,] coordinates { 
(1, 0.536492964)	+-=(0, 	0.005253091)
};
\draw[color=red, ultra thin] (axis cs:\pgfkeysvalueof{/pgfplots/xmin},0.536492964)--(axis cs:\pgfkeysvalueof{/pgfplots/xmax},0.536492964);

\addplot[color=black, dashed, mark=triangle, error bars/.cd, y dir=both, y explicit,] coordinates { 
(1, 0.355668154)	+-=(0,	0.005313959)
};
\draw[color=black, dashed, ultra thin] (axis cs:\pgfkeysvalueof{/pgfplots/xmin},0.355668154)--(axis cs:\pgfkeysvalueof{/pgfplots/xmax},0.355668154);

\nextgroupplot[title={Citation}]
\addplot[color=blue,mark=o,error bars/.cd, y dir=both, y explicit] coordinates {
(1,	0.195147601686) +-=(0, 0.00544251043522)
(2,	0.209316279439) +-=(0, 0.006012864509091) 
(3,	0.21474327209) +-=(0, 0.00588007385345) 
(4,	0.207779694271) +-=(0, 0.00623043193454) 
};

\addplot[color=green, mark=diamond, error bars/.cd, y dir=both, y explicit,] coordinates { 
(1, 0.330430836)	+-=(0,	0.004403963)
};
\draw[color=green, ultra thin] (axis cs:\pgfkeysvalueof{/pgfplots/xmin},0.330430836)--(axis cs:\pgfkeysvalueof{/pgfplots/xmax},0.330430836);
\addplot[color=red, mark=square, error bars/.cd, y dir=both, y explicit,] coordinates { 
(1, 0.440594236)	+-=(0,	0.005212909)
};
\draw[color=red, ultra thin] (axis cs:\pgfkeysvalueof{/pgfplots/xmin},0.440594236)--(axis cs:\pgfkeysvalueof{/pgfplots/xmax},0.440594236);

\addplot[color=black, dashed, mark=triangle, error bars/.cd, y dir=both, y explicit,] coordinates { 
(1, 0.303677576)	+-=(0,	0.003)
};
\draw[color=black, dashed, ultra thin] (axis cs:\pgfkeysvalueof{/pgfplots/xmin},0.303677576)--(axis cs:\pgfkeysvalueof{/pgfplots/xmax},0.303677576);

\nextgroupplot[title={Internet}]
 \addplot[color=blue,mark=o,error bars/.cd, y dir=both, y explicit,] coordinates {
(1,	0.26913049486) +-=(0, 0.0116735355985) 
(2,	0.298125049377) +-=(0, 0.0147916597102) 
(3,	0.294002644737) +-=(0, 0.0188041675103) 
(4,	0.315284441118) +-=(0,0.0200570299392)
};

\addplot[color=green, mark=diamond, error bars/.cd, y dir=both, y explicit,] coordinates { 
(1, 0.330430836)	+-=(0,	0.004403963)
};
\draw[color=green, ultra thin] (axis cs:\pgfkeysvalueof{/pgfplots/xmin},0.330430836)--(axis cs:\pgfkeysvalueof{/pgfplots/xmax},0.330430836);
\addplot[color=red, mark=square, error bars/.cd, y dir=both, y explicit,] coordinates { 
(1, 0.341818582)	+-=(0,	0.004515378)
};
\draw[color=red, ultra thin] (axis cs:\pgfkeysvalueof{/pgfplots/xmin},0.341818582)--(axis cs:\pgfkeysvalueof{/pgfplots/xmax},0.341818582);

\addplot[color=black, dashed, mark=triangle, error bars/.cd, y dir=both, y explicit,] coordinates { 
(1, 0.34983195)	+-=(0,	0.004)
};
\draw[color=black, dashed, ultra thin] (axis cs:\pgfkeysvalueof{/pgfplots/xmin},0.34983195)--(axis cs:\pgfkeysvalueof{/pgfplots/xmax},0.34983195);

\nextgroupplot[title={Purchase}]
\addplot[color=blue,mark=o,error bars/.cd, y dir=both, y explicit,] coordinates {
(1,	0.182550238103) +-=(0,0.00742832685256) 
(2,	 0.2067022790323) +-=(0,0.008316177929251) 
(3,	0.199766105816) +-=(0, 0.0080296248337) 
(4,0.20896399656) +-=(0,0.00809540432419) 
};
  
\addplot[color=green, mark=diamond, error bars/.cd, y dir=both, y explicit,] coordinates { 
(1, 0.597919441)	+-=(0, 0.065008417)
};
\draw[color=green, ultra thin] (axis cs:\pgfkeysvalueof{/pgfplots/xmin},0.597919441)--(axis cs:\pgfkeysvalueof{/pgfplots/xmax},0.597919441);
\addplot[color=red, mark=square, error bars/.cd, y dir=both, y explicit,] coordinates { 
(1, 0.521586592)	+-=(0,	0.005584719)
};
\draw[color=red, ultra thin] (axis cs:\pgfkeysvalueof{/pgfplots/xmin},0.521586592)--(axis cs:\pgfkeysvalueof{/pgfplots/xmax},0.521586592);

\addplot[color=black, dashed, mark=triangle, error bars/.cd, y dir=both, y explicit,] coordinates { 
(1, 0.323422236)	+-=(0,	0.004)
};
\draw[color=black, dashed, ultra thin] (axis cs:\pgfkeysvalueof{/pgfplots/xmin},0.323422236)--(axis cs:\pgfkeysvalueof{/pgfplots/xmax},0.323422236);

\nextgroupplot[title={Wikipedia}]
\addplot[color=blue, mark=o, error bars/.cd, y dir=both, y explicit,] coordinates {
(1,0.342983171368) +-=(0, 0.00769592137092) 
(2,	0.340769050465) +-=(0, 0.00819696899266) 
(3,	0.358552363996) +-=(0,0.009670519836217) 
(4,	0.353278096148) +=(0,0.0101699167079) 
};
\addplot[color=green, mark=diamond, error bars/.cd, y dir=both, y explicit,] coordinates { 
(1, 0.360673596)	+-=(0, 0.004961035)
};
\draw[color=green, ultra thin] (axis cs:\pgfkeysvalueof{/pgfplots/xmin},0.360673596)--(axis cs:\pgfkeysvalueof{/pgfplots/xmax},0.360673596);
\addplot[color=red, mark=square, error bars/.cd, y dir=both, y explicit,] coordinates { 
(1, 0.496033194)	+-=(0,	0.005092394)
};
\draw[color=red, ultra thin] (axis cs:\pgfkeysvalueof{/pgfplots/xmin},0.496033194)--(axis cs:\pgfkeysvalueof{/pgfplots/xmax},0.496033194);

\addplot[color=black, dashed, mark=triangle, error bars/.cd, y dir=both, y explicit,] coordinates { 
(1, 0.418420802)	+-=(0,	0.004)
};
\draw[color=black, dashed, ultra thin] (axis cs:\pgfkeysvalueof{/pgfplots/xmin},0.418420802)--(axis cs:\pgfkeysvalueof{/pgfplots/xmax},0.418420802);
\nop{
\nextgroupplot[title={StackExchange}]
\addplot[color=blue, mark=o, error bars/.cd, y dir=both, y explicit,] coordinates {

(1,	0.193767803635) +=(0, 0.000812907044719) -=(0,0.000812907044719)
(2,	0.13773413306) +=(0, 0.00321549054894) -=(0,0.00321549054894)
(3,	0.145967392015) +=(0,0.00317854046316) -=(0,0.00317854046316)
(4,	0.109412349714) +=(0,0.00318184111912) -=(0,0.00318184111912)
};
\addplot[color=green, mark=diamond, error bars/.cd, y dir=both, y explicit,] coordinates { 
(1, 0.36617018)	+-=(0, 0.004853776)
(2, 0.36617018)	+-=(0, 0.004853776)
(3, 0.36617018)	+-=(0, 0.004853776)
(4, 0.36617018)	+-=(0, 0.004853776)
};
\addplot[color=red, mark=square, error bars/.cd, y dir=both, y explicit,] coordinates { 
(1, 0.439106274)	+-=(0, 0.00443023)
(2, 0.439106274)	+-=(0, 0.00443023)
(3, 0.439106274)	+-=(0, 0.00443023)
(4, 0.439106274)	+-=(0, 0.00443023)
};
}

\end{groupplot}
   
\end{tikzpicture}
\begin{tikzpicture}
    \begin{customlegend}[ legend columns=-1,
  legend style={
    draw=none,
    column sep=1ex,
  },
  legend entries={laHRG, Kron, Chung-Lu, Direct}]

    \addlegendimage{color=blue,mark=o}
    \addlegendimage{color=red, mark=square}
    \addlegendimage{color=green, mark=diamond}
    \addlegendimage{color=black, dashed, mark=triangle, mark options={solid}}
    \end{customlegend}
\end{tikzpicture}
\caption{HRG models are shown to generate graphs with lower (= better) degree distribution distance to the test graph when compared to other models. Splitting nonterminals ($n \geq 2$) sometimes decreases degree distance but sometimes increases it.}
\label{fig:degree_distribution}
\end{figure}

Figure~\ref{fig:degree_distribution} illustrates the results of the degree distribution distance. Recall that the laHRG is identical to HRG~\cite{aguinaga2016growing} when $n=1$. The Kronecker and Chung-Lu do not have an $n$ parameter, so their plots are flat. All points represent the mean of 500 repetitions; each point contains error bars indicating the 95\% confidence intervals -- although many error bars are too small to be seen.

The laHRG model generates graphs that more closely follow the degree distribution of the test graph than graphs generated by Kronecker and Chung-Lu models. Higher nonterminal splitting, \ie, $n>1$, shows little change on the degree distribution distance.

It is expected that the Direct baseline outperforms all graph models, because the Direct baseline simply compares two graphs generated from the exact same generation process, which rewards an overfit model.

Here, the HRG models predict the test graph's degree distribution better than the Direct baseline does; whether this is because they generalize better, or due to chance, or some other reason, would need further analysis to determine. In any case, nonterminal splitting ($n \geq 2$) has only a slight effect on the model, generally attracting the degree distribution toward the training graph's and away from the test graph's. 

Interestingly, the Direct baseline has similar or better performance than the Kronecker and Chung-Lu methods. It is unlikely that these results can be completely explained by overfitting. Instead, Kronecker or Chung-Lu methods may perform poorly due to underfitting, wherein these models do not model the training graph well enough. More work is needed to understand these results.

\begin{figure}[t]
\centering
\begin{tikzpicture} \small
 \begin{groupplot}[compat=1.5,
    group style = {group size = 2 by 3,
        horizontal sep=0.2cm,
        vertical sep=0.8cm,
        xticklabels at=edge bottom,
        yticklabels at=edge left,
        xlabels at=edge bottom,
        ylabels at=edge left,
        },
    width = 1.5in, height = 1.3in,
    title style={yshift=-0.2cm,font={\strut}},
    ymax=7, ymin = 0,
    ylabel = {GCD},
    xlabel={\# of splits ($n$)},
    xtick ={1,2,3,4},
    xticklabel style={align=center}, xticklabels={{1\\\footnotesize (none)},2,3,4},
    legend style={draw=none, fill=none},
]

\nextgroupplot[title={Barabasi-Albert}]
\addplot[blue, mark=o, error bars/.cd, y dir=both, y explicit,] coordinates {
(1,	1.52645556743) +-=(0, 0.114103455808) 
(2,	1.50233735455) +-=(0, 0.111931572735) 
(3,	1.48204137681) +-=(0, 0.11235542601) 
(4,	1.38119737511) +-=(0,0.109095302361) 
};

\addplot[color=green, mark=diamond, error bars/.cd, y dir=both, y explicit,] coordinates { 
(1, 1.816279595)	+-=(0,	0.062672642)
};
\draw[color=green, ultra thin] (axis cs:\pgfkeysvalueof{/pgfplots/xmin},1.816279595)--(axis cs:\pgfkeysvalueof{/pgfplots/xmax},1.816279595);

\addplot[color=red, mark=square, error bars/.cd, y dir=both, y explicit,] coordinates { 
(1, 2.31203642)	+-=(0,	0.074137265)
};
\draw[color=red, ultra thin] (axis cs:\pgfkeysvalueof{/pgfplots/xmin},2.31203642)--(axis cs:\pgfkeysvalueof{/pgfplots/xmax},2.31203642);

\addplot[color=black, dashed, mark=triangle, error bars/.cd, y dir=both, y explicit,] coordinates { 
(1, 1.684949328)	+-=(0,	0.061515287)
};
\draw[color=black, dashed, ultra thin] (axis cs:\pgfkeysvalueof{/pgfplots/xmin},1.684949328)--(axis cs:\pgfkeysvalueof{/pgfplots/xmax},1.684949328);

\nextgroupplot[title={Watts-Strogatz}]    
\addplot[blue, mark=o, error bars/.cd, y dir=both, y explicit,] coordinates {
(1,	3.02058266663) +-=(0,0.0326464145546) 
(2,	3.02249933349) +-=(0, 0.0344255452255) 
(3,	3.02103563617) +-=(0, 0.0347398053952) 
(4,	3.03545823941) +-=(0, 0.0338200204918) 
};

\addplot[color=green, mark=diamond, error bars/.cd, y dir=both, y explicit,] coordinates { 
(1, 5.025377778)	+-=(0,	0.03954414)
};
\draw[color=green, ultra thin] (axis cs:\pgfkeysvalueof{/pgfplots/xmin},5.025377778)--(axis cs:\pgfkeysvalueof{/pgfplots/xmax},5.025377778);

\addplot[color=red, mark=square, error bars/.cd, y dir=both, y explicit,] coordinates { 
(1,5.035254556)	+-=(0, 	0.036062185)
};
\draw[color=red, ultra thin] (axis cs:\pgfkeysvalueof{/pgfplots/xmin},5.0352545565)--(axis cs:\pgfkeysvalueof{/pgfplots/xmax},5.035254556);

\addplot[color=black, dashed, mark=triangle, error bars/.cd, y dir=both, y explicit,] coordinates { 
(1, 2.510669394)	+-=(0,	0.038340046)
};
\draw[color=black, dashed, ultra thin] (axis cs:\pgfkeysvalueof{/pgfplots/xmin},2.510669394)--(axis cs:\pgfkeysvalueof{/pgfplots/xmax},2.510669394);

\nextgroupplot[title={Citation}]
\addplot[blue, mark=o, error bars/.cd, y dir=both, y explicit,] coordinates {
(1,	2.53651692357) +-=(0, 0.0639480147353) 
(2,	2.51903885022) +-=(0, 0.0596142570749) 
(3,	2.47762232581) +-=(0, 0.0639228244503) 
(4,	2.46317278149) +-=(0, 0.0718806787928) 
};

\addplot[color=green, mark=diamond, error bars/.cd, y dir=both, y explicit,] coordinates { 
(1, 5.843521577)	+-=(0, 0.047420128)
};
\draw[color=green, ultra thin] (axis cs:\pgfkeysvalueof{/pgfplots/xmin},5.843521577)--(axis cs:\pgfkeysvalueof{/pgfplots/xmax},5.843521577);

\addplot[color=red, mark=square, error bars/.cd, y dir=both, y explicit,] coordinates { 
(1, 5.646259803)	+-=(0, 0.050087962)
};
\draw[color=red, ultra thin] (axis cs:\pgfkeysvalueof{/pgfplots/xmin},5.646259803)--(axis cs:\pgfkeysvalueof{/pgfplots/xmax},5.646259803);

\addplot[color=black, dashed, mark=triangle, error bars/.cd, y dir=both, y explicit,] coordinates { 
(1, 2.817956436)	+-=(0,	0.042780905)
};
\draw[color=black, dashed, ultra thin] (axis cs:\pgfkeysvalueof{/pgfplots/xmin},2.817956436)--(axis cs:\pgfkeysvalueof{/pgfplots/xmax},2.817956436);

\nextgroupplot[title={Internet}]
\addplot[blue, mark=o, error bars/.cd, y dir=both, y explicit,] coordinates {
(1,	4.53420569805) +-=(0, 0.066783415758) 
(2,	4.55068699246) +-=(0, 0.064371282499) 
(3,	4.87960000427) +-=(0, 0.0868016643465) 
(4,	 4.9496348951) +-=(0,0.0896929035035) 
};

\addplot[color=green, mark=diamond, error bars/.cd, y dir=both, y explicit,] coordinates { 
(1, 4.48868904)	+-=(0, 0.075175688)
};
\draw[color=green, ultra thin] (axis cs:\pgfkeysvalueof{/pgfplots/xmin},4.48868904)--(axis cs:\pgfkeysvalueof{/pgfplots/xmax},4.48868904);

\addplot[color=red, mark=square, error bars/.cd, y dir=both, y explicit,] coordinates { 
(1, 4.881819411)	+-=(0, 0.063193249)
};
\draw[color=red, ultra thin] (axis cs:\pgfkeysvalueof{/pgfplots/xmin},4.881819411)--(axis cs:\pgfkeysvalueof{/pgfplots/xmax},4.881819411);

\addplot[color=black, dashed, mark=triangle, error bars/.cd, y dir=both, y explicit,] coordinates { 
(1, 4.930588248)	+-=(0,	0.072143257)
};
\draw[color=black, dashed, ultra thin] (axis cs:\pgfkeysvalueof{/pgfplots/xmin},4.930588248)--(axis cs:\pgfkeysvalueof{/pgfplots/xmax},4.930588248);

\nextgroupplot[title={Purchase}]    
\addplot[blue, mark=o, error bars/.cd, y dir=both, y explicit,] coordinates {
(1,	2.15767950343) +-=(0,0.0393804564888) 
(2,	2.16842184136) +-=(0,0.0390608946559) 
(3,	2.20741470121) +-=(0,0.042391590104) 
(4,2.37444565596) +-=(0,0.052709973099) 
};

\addplot[color=green, mark=diamond, error bars/.cd, y dir=both, y explicit,] coordinates { 
(1, 6.152948844)	+-=(0, 0.039814542)
};
\draw[color=green, ultra thin] (axis cs:\pgfkeysvalueof{/pgfplots/xmin},6.152948844)--(axis cs:\pgfkeysvalueof{/pgfplots/xmax},6.152948844);

\addplot[color=red, mark=square, error bars/.cd, y dir=both, y explicit,] coordinates { 
(1, 6.100354906)	+-=(0, 0.027406766)
};
\draw[color=red, ultra thin] (axis cs:\pgfkeysvalueof{/pgfplots/xmin},6.100354906)--(axis cs:\pgfkeysvalueof{/pgfplots/xmax},6.100354906);

\addplot[color=black, dashed, mark=triangle, error bars/.cd, y dir=both, y explicit,] coordinates { 
(1, 3.05352128)	+-=(0,	0.036072111)
};
\draw[color=black, dashed, ultra thin] (axis cs:\pgfkeysvalueof{/pgfplots/xmin},3.05352128)--(axis cs:\pgfkeysvalueof{/pgfplots/xmax},3.05352128);

\nextgroupplot[title={Wikipedia}]
\addplot[blue, mark=o, error bars/.cd, y dir=both, y explicit,] coordinates {
(1,	 5.62714872968) +-=(0, 0.0493817415558) 
(2,	5.65185467939) +-=(0, 0.0484799342271) 
(3,	5.76655998876) +-=(0,0.0482848052843) 
(4,	5.83763310851) +-=(0,0.049560152731) 
};

\addplot[color=green, mark=diamond, error bars/.cd, y dir=both, y explicit,] coordinates { 
(1, 2.695849079)	+-=(0,	0.04291834)
};
\draw[color=green, ultra thin] (axis cs:\pgfkeysvalueof{/pgfplots/xmin},2.695849079)--(axis cs:\pgfkeysvalueof{/pgfplots/xmax},2.695849079);

\addplot[color=red, mark=square, error bars/.cd, y dir=both, y explicit,] coordinates { 
(1, 6.895749192)	+-=(0, 0.027406766)
};
\draw[color=red, ultra thin] (axis cs:\pgfkeysvalueof{/pgfplots/xmin},6.895749192)--(axis cs:\pgfkeysvalueof{/pgfplots/xmax},6.895749192);

\addplot[color=black, dashed, mark=triangle, error bars/.cd, y dir=both, y explicit,] coordinates { 
(1, 6.133209338)	+-=(0,	0.071313959)
};
\draw[color=black, dashed, ultra thin] (axis cs:\pgfkeysvalueof{/pgfplots/xmin},6.133209338)--(axis cs:\pgfkeysvalueof{/pgfplots/xmax},6.133209338);

\nop{
\nextgroupplot[title={StackExchange}]
\addplot[blue, mark=o, error bars/.cd, y dir=both, y explicit,] coordinates {
(1,	4.45048016679) +=(0, 0.0313827916246) -=(0,0.0313827916246)
(2,	4.1940203959) +=(0, 0.02316067546119) -=(0,0.0231606754611)
(3,	4.092248985) +=(0,0.035357605302) -=(0,0.035357605302)
(4,	4.0358599857) +=(0,0.022530869585) -=(0,0.022530869585)
};
\addplot[color=green, mark=diamond, error bars/.cd, y dir=both, y explicit,] coordinates { 
(1, 5.121931361)	+-=(0,0.06938329)
(2, 5.121931361)	+-=(0,0.06938329)
(3, 5.121931361)	+-=(0,0.06938329)
(4, 5.121931361)	+-=(0,0.06938329)
};
\addplot[color=red, mark=square, error bars/.cd, y dir=both, y explicit,] coordinates { 
(1, 5.97919441)	+-=(0,0.065008417)
(2, 5.97919441)	+-=(0,0.065008417)
(3, 5.97919441)	+-=(0,0.065008417)
(4, 5.97919441)	+-=(0,0.065008417)
};
}

\end{groupplot}

\end{tikzpicture}
\begin{tikzpicture}
    \begin{customlegend}[ legend columns=-1,
  legend style={
    draw=none,
    column sep=1ex,
  },
  legend entries={laHRG, Kron, Chung-Lu, Direct}]

    \addlegendimage{color=blue,mark=o}
    \addlegendimage{color=red, mark=square}
    \addlegendimage{color=green, mark=diamond}
    \addlegendimage{color=black, dashed, mark=triangle, mark options={solid}}
    \end{customlegend}
\end{tikzpicture}
\caption{HRG models are shown to generate graphs with lower (= better) graphlet correlation difference (GCD) to the test graph, when compared with other models. Splitting nonterminals ($n \geq 2$) sometimes inproves GCD and sometimes decreases it.} 
\label{gcd_real}
\end{figure}
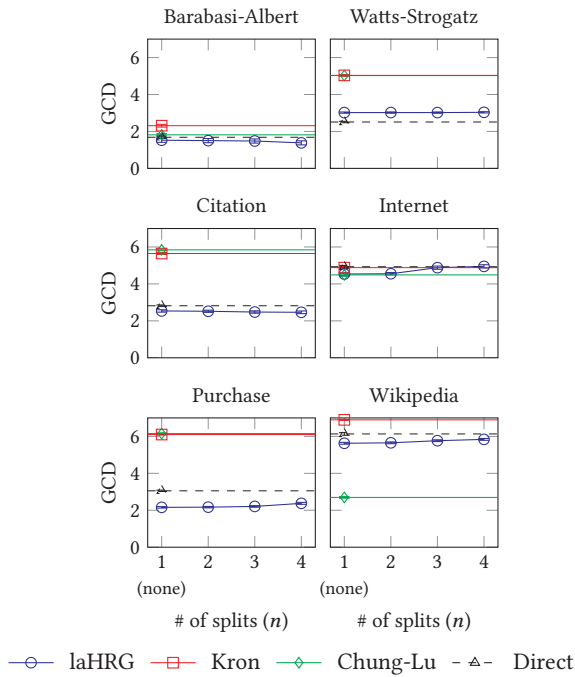

\subsubsection{Graphlet Correlation Distance}

Although the degree distribution is the most well known and widely adopted graph comparison metric, it is far from complete. The degree distribution can be easily mimicked by two very large and different networks. 
For example, previous work has shown that it is easy to construct two or more networks with exactly the same degree distribution but substantially different structure and function~\cite{prvzulj2004modeling,li2005towards}. 
There is mounting evidence which argues that the graphlet comparisons are a better way to measure the similarity between two graphs~\mbox{\cite{prvzulj2007biological,ugander2013subgraph}}. 
Recent work from systems biology has identified a metric called the Graphlet Correlation Distance (GCD). The GCD computes the distance between two graphlet correlation matrices -- one matrix for each graph~\cite{yaveroglu2015proper}. 
Because the GCD is a distance metric, low values indicate high similarity where the GCD is 0 iff the two graphs are isomorphic. 

Figure~\ref{gcd_real} illustrates the results of the GCD. Recall that the laHRG is identical to HRG [10] when n = 1. The Kronecker and Chung-Lu do not have an n parameter, so their plots are flat. All points represent the mean of 500 repetitions; each point contains error bars indicating the 95 confidence intervals – although many error bars are too small to be seen.

The Direct baseline illustrates how similar the training and test graphs are. As expected, we find that the Direct comparison is best on the random Watts-Strogatz graphs. But laHRG outperforms it on all of the real-world graphs.


\subsection{Comparison with Log Likelihood Metric}
GCD and Degree Distribution metrics indicate that laHRG is a better graph generator than other options like Kron and Chung-Lu graph generators, but our experiments seem to suggest that splitting nonterminals in HRG does not have much effect in terms of GCD and Degree Distribution. However, nonterminal splitting does increase log likelihood of the test graph, as explained in Section~\ref{sec:loglikelihood}. This discrepancy is probably because log likelihood metric is able to capture more general structure and properties of a graph than GCD and Degree Distribution. Both GCD and Degree Distribution only focus on a specific graph property, which might not be perfectly correlated with overall structure of a graph. On the other hand, the log likelihood metric we propose does not overemphasize a particular graph property, but directly measures the ability of the graph generator to generate the test graph. 

\begin{table}[t]
\caption{Number of rules/parameters in grammars extracted from training graphs}
\label{fig:sizes}
\begin{center}
\begin{tabular}{lrrrr}
\toprule
& \multicolumn{4}{c}{$n$} \\
\addlinespace[1ex]
Train & \multicolumn{1}{c}{1} & \multicolumn{1}{c}{2} & \multicolumn{1}{c}{3} & \multicolumn{1}{c}{4} \\
\midrule
Citation & 1,156 & 7,193 & 19,885 & 44,410 \\
Internet & 1,005 & 5,686 & 14,057 & 29,247 \\
Purchase & 969 & 6,196 & 18,237 & 38,186 \\
Wikipedia & 1,065 & 6,891 & 20,891 & 42,841 \\
\midrule
Barabasi-Albert & 48 & 298 & 930 & 2,126 \\
Watts-Strogatz & 60 & 346 & 1,023 & 2,380 \\
\bottomrule
\end{tabular}
\end{center}
\end{table}

\section{Grammar Analysis}

Recall that the HRG models merges two production rules if they are identical. Splitting rules produces subrules that have the same structure, but different symbols, so they cannot be merged; splitting nonterminal nodes will therefore increase the size of the grammar. In the worst case, the blowup could be cubic in $n$. Table~\ref{fig:sizes} shows the sizes of all the grammars used in the experiments. Because rules with probability zero are excluded, the blowup is slightly less than cubic.

Here we see a trade-off between model size and performance. The larger the grammar gets, the better it is able to fit the training graph. On the other hand, we prefer smaller models to mitigate the possibility of overfitting. If we had not used separate training and test graphs, it would not be clear how to manage this trade-off, but our evaluation is able to demonstrate that larger grammars (up to a point) are indeed able to generalize to new data.

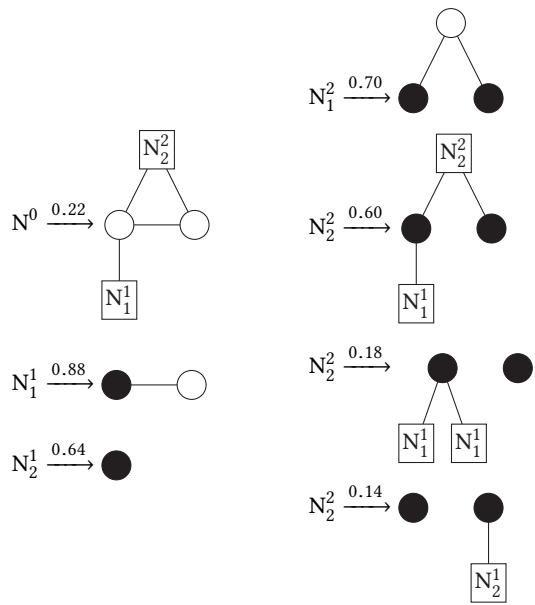
\begin{figure}
\[\begin{aligned}
\n0 &\xrightarrow{0.22}
\begin{tikzpicture}[baseline=-2pt]
\node (n0)[externalnode] at (0,0) {};
\node (n1)[externalnode] at (1,0) {};
\node (e0)[nonterminal] at (0,-1) {$\nss11$};
\node (e01)[nonterminal] at (0.5,1) {$\nss22$};
\draw (n0)--(n1);
\draw (e0)--(n0);
\draw (e01)--(n0);
\draw (e01)--(n1);
\end{tikzpicture} \\[3ex]
\n1_1 &\xrightarrow{0.88} \begin{tikzpicture}[baseline=-2pt]
\node (n0)[externalnode] at (1,0) {};
\node (na)[internalnode] at (0,0) {};
\draw (n0)--(na);
\end{tikzpicture} \\[3ex]
\n1_2 &\xrightarrow{0.64} \begin{tikzpicture}[baseline=-2pt]
\node (na)[internalnode] at (0,0) {};
\end{tikzpicture}
\end{aligned}
\hspace{0.5in}
\begin{aligned}
\n2_1 &\xrightarrow{0.70}
\begin{tikzpicture}[baseline=-2pt]
\node (na)[internalnode] at (0,0) {};
\node (nb)[internalnode] at (1,0) {};
\node (n0)[externalnode] at (0.5,1) {};
\draw (na)--(n0);
\draw (nb)--(n0);
\end{tikzpicture} \\[1ex]
\n2_2 &\xrightarrow{0.60}
\begin{tikzpicture}[baseline=-2pt]
\node (na)[internalnode] at (0,0) {};
\node (nb)[internalnode] at (1,0) {};
\node (eab)[nonterminal] at (0.5,1) {$\nss22$};
\node (ea)[nonterminal] at (0,-1) {$\nss11$};
\draw (eab)--(na);
\draw (eab)--(nb);
\draw (ea)--(na);
\end{tikzpicture} \\[1ex]
\n2_2 &\xrightarrow{0.18}
\begin{tikzpicture}[baseline=-2pt]
\node (na)[internalnode] at (0,0) {};
\node (nb)[internalnode] at (1,0) {};
\node (ea1)[nonterminal] at (-0.35,-1) {$\nss11$};
\node (ea2)[nonterminal] at (0.35,-1) {$\nss11$};
\draw (ea1)--(na);
\draw (ea2)--(na);
\end{tikzpicture} \\[1ex]
\n2_2 &\xrightarrow{0.14}
\begin{tikzpicture}[baseline=-2pt]
\node (na)[internalnode] at (0,0) {};
\node (nb)[internalnode] at (1,0) {};
\node (eb)[nonterminal] at (1,-1) {$\nss12$};
\draw (eb)--(nb);
\end{tikzpicture}
\end{aligned}\]
\caption{Rules extracted from as-topo, 2-split nonterminals, showing only those with probability at least $0.1$ and with at most two external nodes.}
\label{fig:as-grammar}
\end{figure}

What do the HRG grammars look like? The models learned by unsupervised methods like EM can often be difficult to interpret, especially when the number of splits is high and the grammar is large. Figure~\ref{fig:as-grammar} shows selected 2-split rules extracted from the as-topo training graph, namely, those with probability at least 0.1 and with at most two external nodes. 

We can see that the subsymbols behave quite differently from each other. 
For example, the $\n2_1$ rule adds a connection between its two external nodes (via a third node), whereas none of the $\n2_2$ rules adds a connection (perhaps because, as can be seen in the RHS of the $\n0$ rule, they are already neighbors).

What can we learn from these graph grammars? This is an open question. If we assume that the tree decomposition provides a meaningful representation of the original graph, then we may be able to interrogate and assign meaning to these rules depending on their context.  But we save this as a matter for future work.

\section{Conclusion}

This present work identifies and addresses two problems in applying Hyperedge Replacement Grammars (HRGs) to network data~\cite{aguinaga2016growing} by adding latent variables in order to make production rules more sensitive to context and by introducing a principled evaluation methodology that computes the log likelihood that an HRG model generates a graph.

To guard against the possibility of the new model overfitting the original graph, we enforced a separation between the original graph from which the model is trained and a different graph on which the model is tested. This methodology should be better at selecting models that generalize well to new data. We confirmed Aguinaga et al.'s original finding that HRGs perform better than the widely-used Kronecker and Chung-Lu models, and showed that adding latent variables usually improves performance further. 

Furthermore, we evaluated our method against the original HRG model by directly measuring the log-likelihood of the test graphs under all models. This metric is more principled than aggregation of statistics of select graph properties. Under this metric, our method improves over the original in all cases, peaking at either $n=2$ or $3$ splits.

HRGs extracted from tree decompositions are large. 
Splitting nonterminals grows the model even more. But our finding that 2- or 3-split grammars still generalize better to unseen graphs suggests that these models are not unreasonably large.

It remains for future work to test this claim by evaluating other generative graph models on test graphs distinct from training graphs. It should also be possible to simplify the HRG and laHRG models by trying to prune low-probability rules while maintaining high performance. Finally, more analysis is needed to provide an interpretation for the patterns automatically discovered by laHRGs.

\bibliographystyle{ACM-Reference-Format}
\bibliography{hrg}

\end{document}